\documentclass[pre]{revtex4}
\usepackage{amsmath}
\usepackage{graphicx}
\begin{document}
\title{Anomalous Fluctuations in Autoregressive Models with Long-Term Memory}
\author{Hidetsugu Sakaguchi and Haruo Honjo}
\affiliation{Department of Applied Science for Electronics and Materials,
Interdisciplinary Graduate School of Engineering Sciences, Kyushu
University, Kasuga, Fukuoka 816-8580, Japan}
\begin{abstract}
An autoregressive model with a power-law type memory kernel is studied as a stochastic process that exhibits a self-affine-fractal-like behavior for a small time scale. We find numerically that the root-mean-square displacement $\Delta(m)$ for the time interval $m$ increases with a power law as $m^{\alpha}$ with $\alpha<1/2$ for small $m$ but saturates at sufficiently large $m$. The exponent $\alpha$ changes with the power exponent of the memory kernel.
\end{abstract}
\maketitle
\section{Introduction}
The Langevin equation is a typical stochastic differential equation used in statistical physics. Stochastic difference equations are also used  to describe random processes that change stepwise by a unit of day, month, or year. The changes in daily maximum temperature and  stock prices  are examples of such random processes. The autoregressive (AR) model is a simple stochastic difference equation that has been used in statistics and signal processing~\cite{rf:1}. 
The AR model is expressed as 
\begin{equation}
x_n=\sum_{j=1}^ka_jx_{n-j}+\xi_n,
\end{equation}
where $x_n$ is a random variable, $a_j$'s ($j=1,2,\cdots,k$) are parameters that represent the memory effect, and $\xi_n$ is assumed to be a Gaussian white random noise. In equilibrium statistical physics, the memory kernel and  random noises are closely related by the fluctuation-dissipation theorem. However, such a relation is not usually assumed in the AR model, because thermal equilibrium is not considered.   The AR models are simple linear models and used in various fields such as weather prediction and economic forecast. In the AR model, the time correlation function decays exponentially or is a sum of exponential functions. That is, the time correlation decays as $C(m)=\sum_jc_j\lambda_j^m=\sum_j c_j\exp\{\ln(\lambda_j)m\}$, where $c_j$'s are expansion coefficients and  $\lambda_j$'s are eigenvalues of the deterministic linear difference equation~\cite{rf:1,rf:2}:
\begin{equation}
x_n=\sum_{j=1}^ka_jx_{n-j}. 
\end{equation}

There are many linear difference models related to the AR model. The moving average (MA) model is expressed as 
\[x_n=\sum_{j=1}^lb_j\xi_{n-j}+\xi_{n}.\] 
A memory effect is included in the noised term. 
The autoregressive moving average  (ARMA) model is a composite of the AR and MA models. The ARMA $(k,l)$ model is expressed as
\[x_n=\sum_{j=1}^ka_jx_{n-j}+\sum_{j=1}^lb_j\xi_{n-j}+\xi_{n}.\] 
The difference in $x_n$ can be expressed as 
\[x_{n}-x_{n-1}=(1-B)x_n\]
using the lag operator $B$. 
The difference in rank $d$ is defined as $(1-B)^dx_n$.
If two operators are defined as 
\[\phi(B)=1-\sum_{j=1}^ka_jB^j,\;\pi(B)=1+\sum_{j=1}^{l}b_jB^j,\]
the ARMA $(k,l)$ model can be expressed as
\[\phi(B)x_n=\pi(B)\xi_n.\]
The models using $p,q$ of $0,1,2$ or 3 are used in most applications for fluctuating data.   
\begin{figure}
\begin{center}
\includegraphics[height=3.5cm]{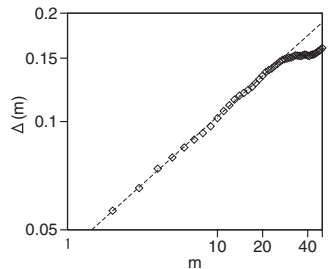}
\end{center}
\caption{Relationship between $m$ and $\Delta(m)$ for the government approval rating in England. The dashed line is a line of $\Delta(m)\propto m^{\alpha}$ with $\alpha=0.375$. }
\label{f1}
\end{figure}

On the other hand, various types of fluctuating time sequence have been studied by many authors from the viewpoint of fractal physics~\cite{rf:3, rf:4}. 
Recently, Honjo et al. have found a certain self-affinity in temporal fluctuations of the monthly government approval ratings of several countries~\cite{rf:5}. Figure 1 shows $\Delta(m)=\langle (x_{n+m}-x_n)^2\rangle^{1/2}$ as a function of $m$, where $0\le x_n\le 1$ is the government approval rating in England between 1979 and 2013.  (Original data are shown using percentage and $x_n$ is the value obtained by diving the data by 100.) $\Delta(m)$ changes as $\Delta(m)\propto m^{\alpha}$ with $\alpha=0.375$ for $m<25$.  However, $\Delta(m)$ tends to saturate for large $m$. In general, the self-affine fractal with $\alpha<1/2$ implies negative correlation. Many examples of self-affine fractals with positive correlation  were reported in random interface problems~\cite{rf:6,rf:7,rf:8,rf:9}. In the normal diffusion, the root-mean-square displacement satisfies $\langle (\Delta x)^2\rangle^{1/2}\sim t^{\alpha}$ with $\alpha=1/2$. The diffusion satisfying $\alpha\ne 1/2$ is called anomalous diffusion. The diffusion in random media sometimes exhibits anomalous diffusion. The exponent $\alpha\ne 1/2$ can be explicitly calculated for diffusions in some fractals~\cite{rf:10}.   
Several mathematical models were constructed to generate self-affine fractals with $\alpha\ne 1/2$. The fractional Brownian motion is a typical example~\cite{rf:3}. 

Some models related to the AR model can also describe random walks. 
The autoregressive integrated moving average (ARIMA) $(k,d,l)$ model is defined using the difference operator $B$ as 
\[\phi(B)(1-B)^dx_n=\pi(B)\xi_n.\]
In the simplest random walk model, the time evolution equation is expressed as $x_n=x_{n-1}+\xi_{n}$. This is interpreted as an ARIMA model of $p=0,d=1$, and $q=0$. The operator $(1-B)^d$ for integer $d$ can be expressed by the binomial expansion as
\[(1-B)^d=\sum_{k=0}^d\frac{d!}{k!(d-k)!}(-B)^k.\] 
If the rank number $d$ is generalized to a non-integer, the model is called the autoregressive fractionally integrated moving average (ARFIMA) model~\cite{rf:11, rf:12,rf:13}.
For the non-integer $d$, the binomial expansion is expressed using the gamma function as    
\[(1-B)^d=\sum_{k=0}^d\frac{\Gamma(d+1)}{\Gamma(k+1)\Gamma(d-k+1)}(-B)^k.\]
In the ARIFIMA model, the root-mean-square displacement $\Delta(m)$ obeys a power law $\Delta(m)\sim m^{\alpha}$ with $\alpha\ne 1/2$ for any $m$, and diverges for $m\rightarrow \infty$. 

In the problem of the government approval rating, $\Delta(m)$ needs to saturate for large $m$, because $x_n$ changes between 0 and 1.   We would like to consider such a random process in which $\Delta(m)\propto  m^{\alpha}$ for small $m$ but saturates for large $m$.  Because $\{\Delta(m)\}^2=\langle (x_{n+m}-x_n)^2\rangle=2\langle x_n^2\rangle -2\langle x_{n+m}x_n\rangle$, the time correlation $C(m)=\langle x_{n+m}x_n\rangle$ decays as $C(m)=C(0)(1-\beta m^{2\alpha})$. This decay law can be applied only for the time range where $\{\Delta(m)\}^2\propto m^{2\alpha}$ is satisfied. When $m$ becomes large, $\{\Delta(m)\}^2$ saturates toward a finite value and $C(m)$ decays to 0. 
We do not yet understand the reason why government approval ratings exhibit such anomalous fluctuation, but the existence of obstinate persons with long-term memory might be a reason for the anomalous fluctuation. Motivated by the anomalous fluctuations, we will study an abstract AR model with a long-term memory kernel in this paper. We study the simple AR model to reproduce the power-law behavior of $\Delta(m)$ for small $m$. Our simple AR model is different from the ARFIMA model, in the sense that the difference operator $B$ of the fractional rank $d$ is not used, and $\Delta(m)$ saturates at large $m$.
\section{Autoregressive Model with Long-Term Memory}
As an AR model with long-term memory, we assume that  $a_i$'s decay in a power law. The model equation is written as
\begin{equation}
x_n=A\sum_{j=1}^{\infty}\frac{x_{n-j}}{j^q}+\xi_n,
\end{equation}
where $A=\gamma/\zeta_q$ and $\zeta_q=\sum_{j=1}^{\infty}(1/j^q)$.  $\zeta_q$ is the Riemann zeta function and it converges for $q>1$. That is, $a_j$ in Eq.~(1) is set to be $A/j^q=\gamma/(\zeta_qj^q)$. For $\gamma<1$, the AR model Eq.~(3) generates a stochastic process where the variance $\langle x_n^2\rangle$ is finite. 
At $\gamma=1$, the variance $\langle x_n^2\rangle$ increases with $n$ indefinitely like Brownian motions. This is because the deterministic linear difference equation without the last noise term, 
\[
x_n=A\sum_{j=1}^{\infty}\frac{x_{n-j}}{j^q}
\]
has an eigenvalue 1 at $\gamma=1$, and $x_n$ is not attracted to $x=0$.   

Because the AR model is a linear equation, the solution can be obtained by the Fourier transform method. We consider a time series of length $N$.  
From Eq.~(3), the Fourier transform $x(\omega_k)=\sum_{n=0}^{N-1}x_ne^{-i\omega_k n}$ for $\omega_k=2\pi k/N$  satisfies  
\[
x(\omega_k)=F(\omega_k)x(\omega_k)+\xi(\omega_k),
\]
where $F(\omega_k)=\sum_{j=1}^{N-1}(A/j^{q})e^{-i\omega_k j}$  and $\xi(\omega_k)=\sum_{j=0}^{N-1}\xi_ne^{-i\omega_k j}$.  $x(\omega_k)$ is therefore solved as \begin{equation}
x(\omega_k)=\frac{\xi(\omega_k)}{1-F(\omega_k)}.
\end{equation} 
The mean square displacement $\{\Delta(m)\}^2$ is evaluated as
\begin{equation}
\{\Delta(m)\}^2=\frac{4}{N}\sum_{k=0}^{N/2-1}\langle |x(\omega_k)|^2\rangle \{1-\cos(\omega_k m)\}=\frac{4\sigma^2}{N}\sum_{k=0}^{N/2-1}\frac{1-\cos(\omega_km)}{|1-F(k)|^2},
\end{equation}
where $\sigma^2=\langle \xi_n^2\rangle$ denotes the strength of random noises. 

Figure 2(a) is an example of a random time sequence at $q=2$, $\gamma=0.98$, and $\sigma=0.5$. 
The solid lines in Fig.~2(b) show the relationship between $m$ and $\Delta(m)$ obtained by the direct numerical simulation of Eq.~(3) for $q=2$ and 2.5 at $\gamma=0.98$ and $\sigma=0.5$.
\begin{figure}
\begin{center}
\includegraphics[height=4.cm]{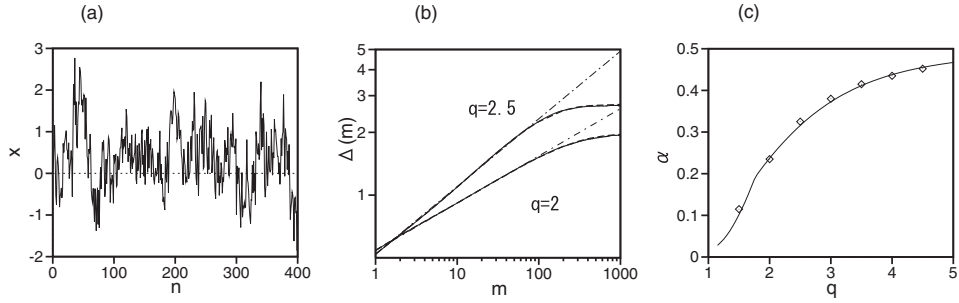}
\end{center}
\caption{(a) Time sequence of $x_n$ for $q=2,\beta=0.98$, and $\sigma=0.5$. 
(b) Solid lines denote the relationship between $m$ and $\Delta(m)$ for the fluctuating time sequences generated by Eq.~(3) for $q=2$ and 2.5  at $\beta=0.98$ and $\sigma=0.5$. Dashed lines denote $\Delta(m)$ calculated using Eq.~(5). However, the difference between the solid and dashed lines is hardly visible owing to the overlap. Two alternate long and short dashed lines denote the lines of $m^{\alpha}$ with $\alpha=0.223$ and 0.325. 
(c) Relationship between $q$ and the exponent $\alpha$ for $\gamma=0.98$. Rhombi denote the values evaluated from direct numerical simulations, and the solid curve denotes the result obtained from Eqs.~(9) and (10).}
\label{f2}
\end{figure}
The dashed lines denote the relationship between $m$ and $\Delta(m)$ obtained by the numerical calculation of Eq.~(5), which is consistent with direct numerical simulation. The difference between the data calculated using Eq.~(5) and the direct numerical simulations is hardly visible in Fig.~2(b). 
$\Delta(m)$ obeys the power law $\Delta(m)\propto m^{\alpha}$ for $m<100$, and tends to saturate for large $m$. The alternate long and short dashed lines are the linear fitting in the double-logarithmic plot. The exponent $\alpha$ is evaluated as 0.223 for $q=2$ and 0.325 for $q=2.5$. The exponent $\alpha$ increases with $q$.  

The relation between the time correlation function $C(m)$ and the parameters $a_j$ is known as the Yule-Walker equations~\cite{rf:14}.
If we assume a power law for the time correlation function, $C(m)=C(0)(1-\beta m^p)$, there are  three unknown parameters: $C(0),\beta$, and $p$. We can estimate the three unknown parameters from the time correlation $C(m)$ at $m=0, 1$, and 2. Taking the average with respect to $\xi$ after multiplying both sides of Eq.~(3) by $x_n$, $C(0)$ satisfies
\begin{equation}
C(0)=\langle x_n^2\rangle=A\sum_{m=1}^{\infty}\frac{C(m)}{m^q}+\sigma^2.
\end{equation}
This is the first Yule-Walker equation for $C(m)$ and $\sigma$.  
Substitution of $C(m)=C(0)(1-\beta m^{p})$ where $p=2\alpha$ into Eq.~(6) yields
\begin{equation}
C(0)=\frac{\sigma^2}{1-\gamma+\gamma\beta \zeta_{q-p}/\zeta_q}.
\end{equation}
Similarly, by taking the average with respect to $\xi$ after multiplying both sides of Eq.~(3) by $x_{n-1}$ and $x_{n-2}$, $C(1)$ and C(2) are expressed as 
\begin{eqnarray} 
C(1)&=&C(0)(1-\beta)=C(0)\{\gamma-A\beta\sum_{j=2}^{\infty}\frac{(j-1)^p}{j^q}\},\nonumber\\
C(2)&=&C(0)(1-\beta 2^p)=C(0)\{-A\beta+\gamma-A\beta\sum_{j=3}^{\infty}\frac{(j-2)^p}{j^q}\}.
\end{eqnarray}    
From the expressions of $C(1)$ and $C(2)$, the following coupled equations are obtained,
\begin{eqnarray}
\beta&=&\frac{1-\gamma}{1-\gamma/\zeta_q\sum_{j=2}^{\infty}(j-1)^p/j^q},\\
1-\gamma&=&\beta \left (2^p-\frac{\gamma}{\zeta_q}\right )-\frac{\gamma}{\zeta_q}\beta\sum_{j=3}^{\infty}\frac{(j-2)^p}{j^q}.
\end{eqnarray}
By the coupled equations Eqs.~(9) and (10), $p=2\alpha$ and $\beta$  are solved at least numerically, and the variance $C(0)$ is calculated using Eq.~(7). 
Rhombi in Fig.~2(c) show the numerically evaluated value of $\alpha$ and the solid curve denotes $\alpha=p/2$ obtained from the solution of the coupled equations Eqs.~(9) and (10) at $\gamma=0.98$.  Fairly good agreement is observed. 

\begin{figure}
\begin{center}
\includegraphics[height=4.cm]{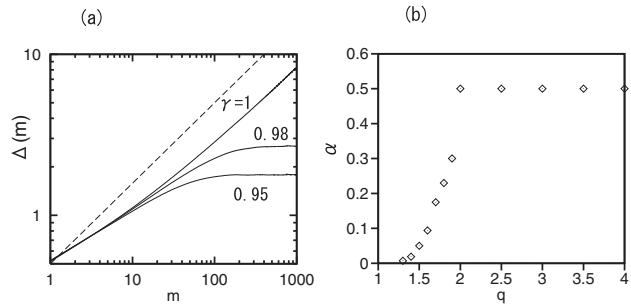}
\end{center}
\caption{(a) Relationship between $m$ and $\Delta(m)$ at $\gamma=0.95,0.98$, and 1 for $q=2.5$ and $\sigma=0.5$ obtained by direct numerical simulations of Eq.~(3). (b) Relationship between $q$ and the exponent $\alpha$ for $\gamma=1$.}
\label{f3}
\end{figure}
\begin{figure}
\begin{center}
\includegraphics[height=3.5cm]{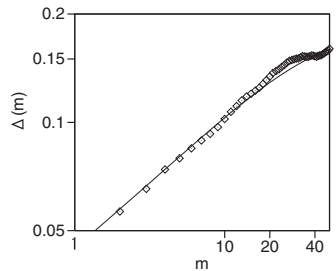}
\end{center}
\caption{Relationship between $m$ and $\Delta(m)$ for the government approval rating in England. The solid line is $\Delta(m)$ for the AR model at  $q=3,\gamma=0.95$, and $\sigma=0.147$.}
\label{f4}
\end{figure}
The parameter $\gamma$ determines the temporal range in which the power law is observed. 
Figure 3(a) shows the relationship between $m$ and $\Delta(m)$ at $\gamma=0.95,0.98$, and 1 for $q=2.5$ and $\sigma=0.5$ obtained by direct numerical simulations of Eq.~(3). Three lines overlap well for $m<20$, and exhibit a power law of exponent $\alpha=0.325$.   
As $\gamma$ decreases, $\Delta(m)$ saturates at small $m$. At $\gamma=1$, $\Delta(m)$ increases indefinitely. However, there is a cross-over in the power law  of $\Delta(m)$, that is,  $\Delta(m)\sim m^{0.325}$ for small $m$, but $\Delta(m)\sim m^{0.5}$ for sufficiently large $m$. The dashed line in Fig.~3(a) is a line of $m^{0.5}$. The exponent $\alpha=1/2$ implies that the statistical property of long-time fluctuations at $q=2.5$ and $\gamma=1$ is equivalent to a simple random walk. We have evaluated the exponent $\alpha$ for sufficiently large $m$ by changing $q$ at $\gamma=1$. Figure 3(b) shows the exponent $\alpha$ as a function of $q$. The exponent $\alpha$ increases rapidly near $q=2$ and becomes $1/2$ for $q> 2$.  

In Fig.~4, we show the data of the government approval rating and $\Delta(m)$ for the AR model (3) at $q=3,\gamma=0.95$, and $\sigma=0.147$. The characteristics of $\Delta(m)\propto m^{\alpha}$ with $\alpha\sim 0.375$ for small $m$ and the saturation for large $m$ are reproduced by the AR model. 
\section{Summary}
We have proposed an AR model with long-term memory obeying a power law. We have found that $\Delta(m)$ increases with a power law of an exponent $\alpha<1/2$ and saturates for large $m$. In Sect. 1, we showed an example of a government approval rating that exhibits such anomalous fluctuations. We would like to investigate other stochastic processes in nature and social physics that exhibit a similar anomalous statistical property in the future. At $\gamma=1$, $\Delta(m)$ increases $m^{\alpha}$ even for infinitely large $m$. The scaling behavior at large $m$ is similar to that of the ARFIMA model. The construction methods of model equations are different; however, some mathematical relationship between the two models might exist, which is left for a future study.


\begin{thebibliography}{99}
\bibitem{rf:1} G.~Box and G.~M.~Jenkins, {\it Time Series Analysis: Forecasting and Control} (Holden-Day, San Francisco, 1970).
\bibitem{rf:2} J.~Gao, J.~Hu, W-W.~Tung, Y.~Cao, N.~Sarshar, and V.~P.~Roychowdhury, Phys. Rev. E {\bf 73}, 016117 (2006). 
\bibitem{rf:3} B.~B.~Mandelbrot, {\it The Fractal Geometry of Nature} (W.~H.~Freeman and Co, San Francisco, 1982)
\bibitem{rf:4} R.~N.~Mantegna and H.~E.~Stanley, Nature {\bf 376}, 46 (1995). 
\bibitem{rf:5} H.~Honjo, M.~Sano, H.~Miki, and H.~Sakaguchi, Physica A {\bf 428}, 266 (2015).
\bibitem{rf:6} T.~Vicsek, M.~Cerzo, and V.~K.~Horv\'ath, Physica A {\bf 167}, 315 (1990).
\bibitem{rf:7} S.~V.~Buldyrev, A.-L.~Bar\'abasi, F.~Caserta, S.~Havlin, H.~E.~Stanley, and T.~Vicsek, Phys. Rev. A {\bf 45}, R8313 (1992).
\bibitem{rf:8} H.~Honjo and S.~Ohta, Phys. Rev. E {\bf 49}, R1808 (1994).
\bibitem{rf:9} H.~Sakaguchi, Phys. Rev. E {\bf 82}, 032101 (2010).
\bibitem{rf:10} S.~Havlin and D.~Ben-Avraham, Adv. Phys. {\bf 36}, 695 (1987).
\bibitem{rf:11} G.~W.~J.~Granger and R.~Joyeaux, J. Time Series Anal. {\bf 1}, 15 (1980).  
\bibitem{rf:12} J.~R.~Hosking, Biometrika {\bf 68}, 165 (1981).
\bibitem{rf:13} P.~S.~Kokoszka and M.~S.~Taqqu, Stochastic Processes Appl. {\bf 60}, 19 (1995).
\bibitem{rf:14} G.~U.~Yule, Philos. Trans. R. Soc. London {\bf 226}, A267 (1927). 
\end{thebibliography}
\end{document}